\documentclass[12pt]{article}
\usepackage{graphicx}
\usepackage{parskip}
\usepackage{ amssymb }
\usepackage{xspace}
\usepackage[export]{adjustbox}
\usepackage{subcaption}

\newcommand{\Pell}{\ensuremath{\mathrm{\ell}}}
\newcommand{\PGn}{\ensuremath{\mathrm{\nu}}}

\newcommand{\PZ}{\ensuremath{\mathrm{Z}}\xspace}

\newcommand{\PH}{\ensuremath{\mathrm{H}}\xspace}

\newcommand{\Pq}{\ensuremath{\mathrm{q}}\xspace}

\newcommand{\ttbar}{\ensuremath{\mathrm{t\bar{t}}}}

\newcommand{\ttH}{\ensuremath{\ttbar\PH}\xspace}

\newcommand{\ttll}{\ensuremath{\ttbar\Pell\overline{\Pell}}\xspace}
\newcommand{\ttlnu}{\ensuremath{\ttbar\Pell\PGn}\xspace}
\newcommand{\tllq}{\ensuremath{\mathrm{t\Pell\overline{\Pell}\Pq}}\xspace}
\newcommand{\tHq}{\ensuremath{\mathrm{t\PH\Pq}}\xspace}
\newcommand{\tttt}{\ensuremath{\mathrm{\ttbar\ttbar}}\xspace}

\newcommand*{\eftOp}[4]{\ensuremath{
    {#4
    \ifx\empty#3\empty\ifx\empty#1\empty\else^{#1}\fi\else^{#1(#3)}\fi
    \ifx\empty#2\empty\else_{#2}\fi}
}}

\newcommand{\Lagr}{\mathcal{L}}
\newcommand{\me}{\mathcal{M}}
\newcommand{\weight}{\textit{w}}

\newcommand{\pt}{p_{\mathrm{T}}}
\newcommand{\ptz}{p_\mathrm{T}(\mathrm{\PZ})}
\newcommand{\ptljz}{p_{\mathrm{T}}(\ell {\textrm{j}})_{\textrm{max}}}

\newcommand{\twolss}{\ensuremath{2\Pell\textrm{ss}}\xspace}

\newcommand{\threel}{\ensuremath{3\Pell}\xspace}

\newcommand{\fourl}{\ensuremath{4\Pell}\xspace}

\newcommand{\twohqV}{\ensuremath{\textrm{2hqV}}\xspace}
\newcommand{\fourhq}{\ensuremath{\textrm{4hq}}\xspace}
\newcommand{\twohqtwolq}{\ensuremath{\textrm{2hq2lq}}\xspace}
\newcommand{\twohqtwolep}{\ensuremath{\textrm{2hq2\Pell}}\xspace}

\newcommand{\ctp}  {\eftOp{}{t\varphi}{}{c}\xspace}

\newcommand{\cpQM} {\eftOp{-}{\varphi Q}{}{c}\xspace}
\newcommand{\cpQa} {\eftOp{3}{\varphi Q}{}{c}\xspace}
\newcommand{\cpt}  {\eftOp{}{\varphi t}{}{c}\xspace}
\newcommand{\cptb} {\eftOp{}{\varphi tb}{}{c}\xspace}

\newcommand{\ctW}  {\eftOp{}{tW}{}{c}\xspace}

\newcommand{\ctZ}  {\eftOp{}{tZ}{}{c}\xspace}

\newcommand{\cbW}  {\eftOp{}{bW}{}{c}\xspace}

\newcommand{\ctG}  {\eftOp{}{t G}{}{c}\xspace}

\newcommand{\cQla} {\eftOp{3}{Q \ell}{\ell}{c}\xspace}
\newcommand{\cQlM} {\eftOp{-}{Q \ell}{\ell}{c}\xspace}
\newcommand{\cQe}  {\eftOp{}{Q e}{\ell}{c}\xspace}
\newcommand{\ctl}  {\eftOp{}{t \ell}{\ell}{c}\xspace}
\newcommand{\cte}  {\eftOp{}{t e}{\ell}{c}\xspace}
\newcommand{\ctlS} {\eftOp{S}{t}{\ell}{c}\xspace}

\newcommand{\ctlT} {\eftOp{T}{t}{\ell}{c}\xspace}

\newcommand{\cttOne}   {\eftOp{1}{t t}{}{c}\xspace}
\newcommand{\cQQOne}   {\eftOp{1}{Q Q}{}{c}\xspace}
\newcommand{\cQtOne}   {\eftOp{1}{Q t}{}{c}\xspace}
\newcommand{\cQtEight} {\eftOp{8}{Q t}{}{c}\xspace}
\newcommand{\cQqOneThree}   {\eftOp{31}{Q q}{}{c}\xspace}
\newcommand{\cQqEightThree} {\eftOp{38}{Q q}{}{c}\xspace}
\newcommand{\cQqOneOne}     {\eftOp{11}{Q q}{}{c}\xspace}
\newcommand{\ctqOne}        {\eftOp{1}{t q}{}{c}\xspace}
\newcommand{\cQqEightOne}   {\eftOp{18}{Q q}{}{c}\xspace}
\newcommand{\ctqEight}      {\eftOp{8}{t q}{}{c}\xspace}

\newcommand\pubnumber{Experiment-1-1}
\newcommand\pubdate{\today}

\def\institute{Department of Physics\\
The Ohio State University\\
191 West Woodruff Ave\\
Columbus, OH, 43210, USA}
\def\authemail{\footnote{Contact: aashwin.basnet@cern.ch}}
\def\Title#1{\begin{center} {\Large #1 } \end{center}}
\def\Author#1{\begin{center}{ \sc #1} \end{center}}
\def\Address#1{\begin{center}{ \it #1} \end{center}}

\newcommand\pubblock{\rightline{\begin{tabular}{l} \pubnumber\\
         \pubdate  \end{tabular}}}
\newenvironment{Abstract}{\begin{quotation}  }{\end{quotation}}
\newenvironment{Presented}{\begin{quotation} \begin{center} 
             PRESENTED AT\end{center}\bigskip 
      \begin{center}\begin{large}}{\end{large}\end{center} \end{quotation}}

\def\beq{\begin{equation}}
\def\eeq#1{\label{#1}\end{equation}}
\def\eeqn{\end{equation}}

\def\beqa{\begin{eqnarray}}
\def\eeqa#1{\label{#1}\end{eqnarray}}
\def\eeqan{\end{eqnarray}}

\let\bar=\overbar

\def\Dslash{\not{\hbox{\kern-4pt $D$}}}
\def\dslash{\not{\hbox{\kern-2pt $\del$}}}

\def\msb{{\bar{\ssstyle M \kern -1pt S}}}

\begin{document}
\begin{titlepage}
\pubblock

\vfill
\Title{Probing EFT models using top quark production in multilepton final states}
\vfill
\Author{ Aashwin Basnet\authemail}
\begin{center}
    On behalf of the CMS Collaboration
\end{center}
\Address{\institute}
\vfill
\begin{Abstract}
Using data consisting of top quarks produced with additional final leptons collected by the CMS detector at a center-of-mass energy of $\sqrt{s}=$13 TeV from 2016 to 2018 (138 fb$^{-1}$), a search for beyond standard model (BSM) physics is presented. The BSM physics is probed in the context of Effective Field Theory (EFT) by parameterizing potential new physics effects in terms of 26 dimension-six EFT operators. The data are categorized based on lepton multiplicity, total lepton charge, jet multiplicities, and b-tagged jet multiplicities. To gain further sensitivity to potential new physics (NP) effects, events in each jet category are binned using kinematic differential distributions. A simultaneous fit to data is performed to put constraints on the 26 operators. The results are consistent with the standard model prediction.
\end{Abstract}
\vfill
\begin{Presented}
$16^\mathrm{th}$ International Workshop on Top Quark Physics\\
(Top2023), 24--29 September, 2023
\end{Presented}
\vfill
\end{titlepage}
\def\thefootnote{\fnsymbol{footnote}}
\setcounter{footnote}{0}

\section{Introduction}

The Standard Model (SM) of particle physics offers our current best understanding of the interactions between the elementary particles and fundamental forces of the nature, but it has several limitations such as its inability to explain phenomena like dark matter, dark energy, gravity, and neutrino masses. This indicates that the SM might not be the ultimate theory of nature - a prime motivation to search for BSM physics. In recent times, experiments dedicated towards direct searches for new physics (NP) have not identified any conclusive signs. But what if the current energy scale at the LHC is simply insufficient to produce new on-shell particles? Hence, it makes sense to adopt some indirect means of probing high energy scales at the LHC, one of which is Standard Model Effective Field Theory (SMEFT). SMEFT is a powerful relatively model-independent framework that allows us to parameterize potential deviations from the SM predictions. 

The key assumption of SMEFT is that the SM is correct and a complete theory at the energy scale that can be currently probed. Furthermore, the NP is characterized by some unknown energy scale $\Lambda > \Lambda_{\mathrm{LHC}}$. The SM Lagrangian is then expanded by higher dimensional operators ($d > 4$), which are purely made of SM fields and their derivatives and characterize the NP effects. The strength of this interaction is given by dimensionless parameters called Wilson Coefficients (WCs). The effective lagrangian can be written as
\begin{equation}
    \Lagr_{\mathrm{EFT}}=\Lagr_{\mathrm{SM}}+\sum_{i,d>4} \frac{c_{i}^{(d)}}{\Lambda^{d-4}}\mathcal{O}_{i}^{(d)},
\end{equation}
where $\Lagr_{SM}$ is the SM lagrangian, $c_i$ is an $\textit{i}$th WC, $\mathcal{O}_{i}$ is an $\textit{i}$th effective operator, and $d$ is the mass dimension. Since odd-dimension operators violate lepton and baryon number conservation, the leading order contribution arises from the $d=$ 6 terms and are considered in this analysis \cite{Degrande_2013}.

The analysis described in this proceeding uses data involving associated top quark production in multilepton final states, collected by the CMS experiment \cite{cmsdetector} from 2016 to 2018 (integrated luminosity of 138 fb$^{-1}$). We model EFT effects directly at the detector level and perform a simultaneous global fitting of 26 EFT operators that strongly impact six relatively rare signal processes. The signal processes considered are \ttH, \ttlnu, \ttll, \tllq, \tHq, and \tttt. Full details of the analysis can be found in Ref. \cite{top22006}.

\section{Event categorization}
Events are categorized into two leptons same-sign (\twolss), three leptons (\threel), and four or more leptons (\fourl). Each of the lepton categories are further subdivided based on jet multiplicity, b-tagged jet multiplicity, lepton charge sum, and whether or not the invariant mass of same-flavor opposite-sign (SFOS) lepton pair is close to Z boson mass. Overall, this leads to 43 distinct event categories. This categorization aids in improving the sensitivity of the analysis by allowing us to distinguish the effects of the WCs on several signal processes.

To gain additional sensitivity to EFT effects, events in each of the 43 categories are binned using two kinematic variables. They are: i) $\ptz$, defined as the $p_T$ of the SFOS lepton pair, and ii) $\ptljz$, defined as:
\begin{equation}
    \ptljz = \mathrm{max}\left( \mathrm{max} \left(\pt(\ell,\ell') \right),\, \mathrm{max} \left(\pt(j,j')\right),\,\mathrm{max} \left(\pt(\ell,j)\right)\right),
\end{equation}
where $\ell$ and $j$ are leptons and jets, respectively, and $\mathrm{max} \left( \pt(\ell,\ell')\right)$ indicates the $\pt$ of the pair of unique leptons with the largest $\pt$, and same for the other terms. We use $\ptz$ for the 3$\ell$-onZ categories (except for on-Z 2b 2j/3j categories), and $\ptljz$ for the remaining categories. For fitting purpose, we use 4 bins in $\ptljz$ and 5 bins in $\ptz$, which leads to 178 analysis bins in total. This differential kinematic binning helped us improve the sensitivity of the analysis by about a factor of 2 as compared to the inclusive jet multiplicity binning. 

\section{EFT operators and parametrization}
This analysis includes all operators from the dim6top model with significant impact on the six signal processes, which leads to 26 operators \cite{dim6topnote}. These operators fall into four main categories: i) Two-heavy with bosons (\twohqV): \ctp, \cpQM, \cpQa, \cpt, \cptb, \ctW, \ctZ, \cbW, \ctG, ii) Two-heavy-two-lepton (\twohqtwolep): \cQla, \cQlM, \cQe, \ctl, \cte, \ctlS, \ctlT, iii) Two-heavy-two-light (\twohqtwolq): \cQqOneThree, \cQqEightThree, \cQqOneOne, \cQqEightOne, \ctqOne, \ctqEight, and iv) Four-heavy (\fourhq): \cQQOne, \cQtOne, \cQtEight, \cttOne. 

To detect effects of NP on the observed yields, the predicted yields must be parameterized in terms of the WCs. This analysis uses the tree-level EFT modeling presented in Ref. \cite{dim6topnote}. Starting at the Matrix Element (ME) level, we can write: 
\begin{equation}
    \me = \me_{\mathrm{SM}}+\sum_{i}\frac{c_i}{\Lambda^2}\me_{i},
\end{equation}
where $\me_{\mathrm{SM}}$ is the SM ME and $\me_{i}$ are the MEs corresponding to NP components. The cross-section - inclusive or differential - is directly to proportional ME squared, which means that the cross-section is a quadratic function of the WCs. Since each event's weight $\weight$ is its contribution to the inclusive cross-section, we can write
\begin{equation}
    \weight_i = \mathrm{s_{\mathrm{0i}}}+\sum_{j}\mathrm{s_{\mathrm{1ij}}}c_j+\sum_{j}\mathrm{s_{\mathrm{2ij}}}c_j^{2}+\sum_{j\neq k}\mathrm{s_{\mathrm{3ijk}}}c_jc_k,
\label{xsec_param}
\end{equation}
where $\mathrm{s_0}$, $\mathrm{s_1}$, $\mathrm{s_2}$, $\mathrm{s_3}$ are the structure constants representing pure SM, interference between NP and SM, pure NP, and interference between NP contributions, respectively. Since this analysis considers 26 WCs, the event weight is a 26 dimensional quadratic function of the WCs. We generate simulations with EFT effects modeled at tree level at a certain starting point in WC phase space and use MadGraph's event reweighting technique to evaluate the weights in at least 378 alternative points \cite{Mattelaer_2016}. This allows us to obtain the full EFT parametrization of the event weight.   

For any observable bin in the analysis, the predicted yield is given by the sum of event weights of the events that pass the selection criteria. Hence, the predicted yield for any given analysis bin is also quadratically parameterized in terms of the WCs. Further details on EFT parametrization of the event yields can be found in Ref. \cite{top19001}.
\begin{figure}[htp!]
\centering
\includegraphics[width=1\textwidth]{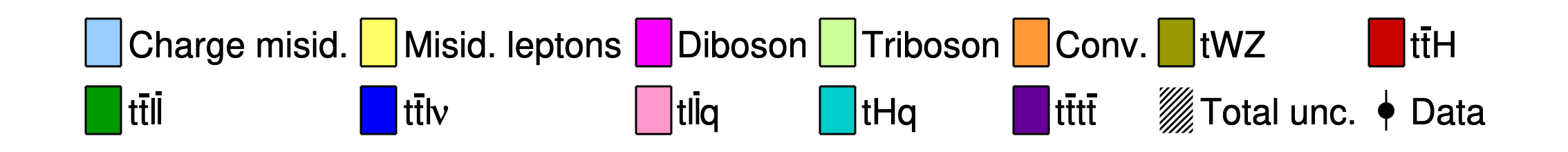}
\includegraphics[width=1\textwidth]{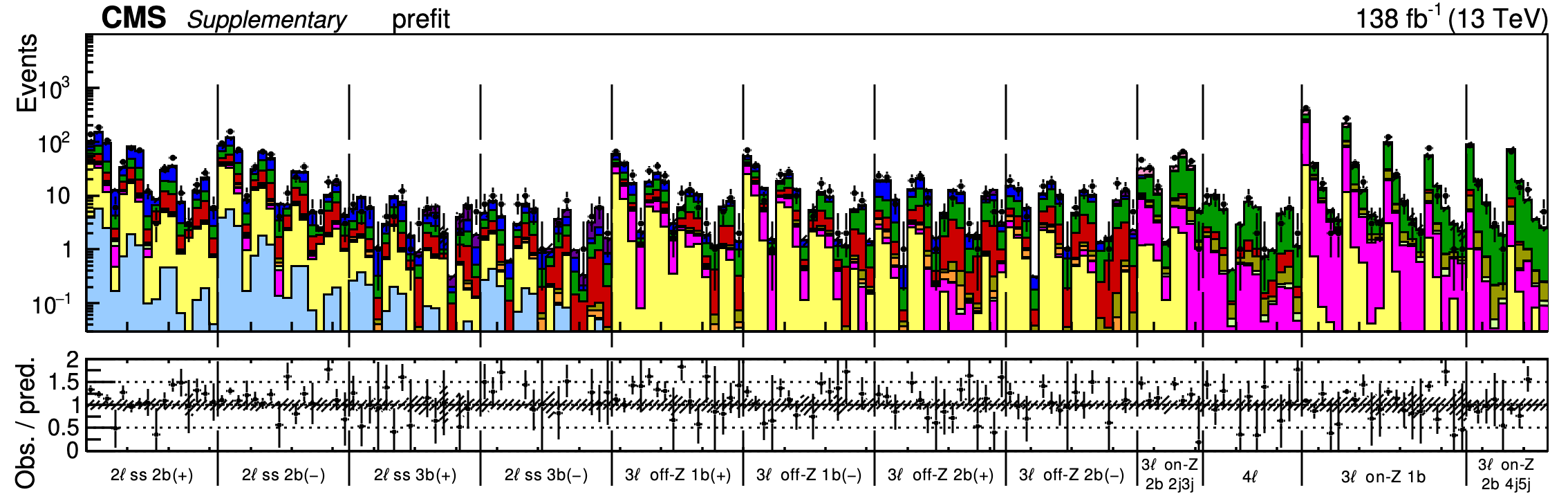}
\caption{Pre-fit plot showing predicted yields with EFT parametrization (colorful histograms) and observation (black dots). \cite{top22006}}
\label{fig:megaprefit_plot}
\end{figure}

In the pre-fit plot in Figure~\ref{fig:megaprefit_plot}, all the bins have been reweighted to SM scenario i.e. all WCs set to 0. Statistical inference is then performed on this histogram, where we can turn the knobs on these 26 WCs simultaneously. This changes the overall shape and normalization across all 178 analysis bins. Using profiled likelihood fitting technique, we are then able to extract the best-fit values and the corresponding uncertainties for the WCs which give the best agreement between the predicted yields and observation. 

\section{Results}
We have extracted 1$\sigma$ and 2$\sigma$ Confidence Intervals (CIs) for the 26 WCs under two scenarios: i) scanning a WC while freezing the other 25 to SM value of zero and ii) scanning a WC while profiling the other 25 WCs. 
\begin{figure}[!htp]
	\begin{minipage}{0.55\textwidth}
		\includegraphics[width=1\textwidth]{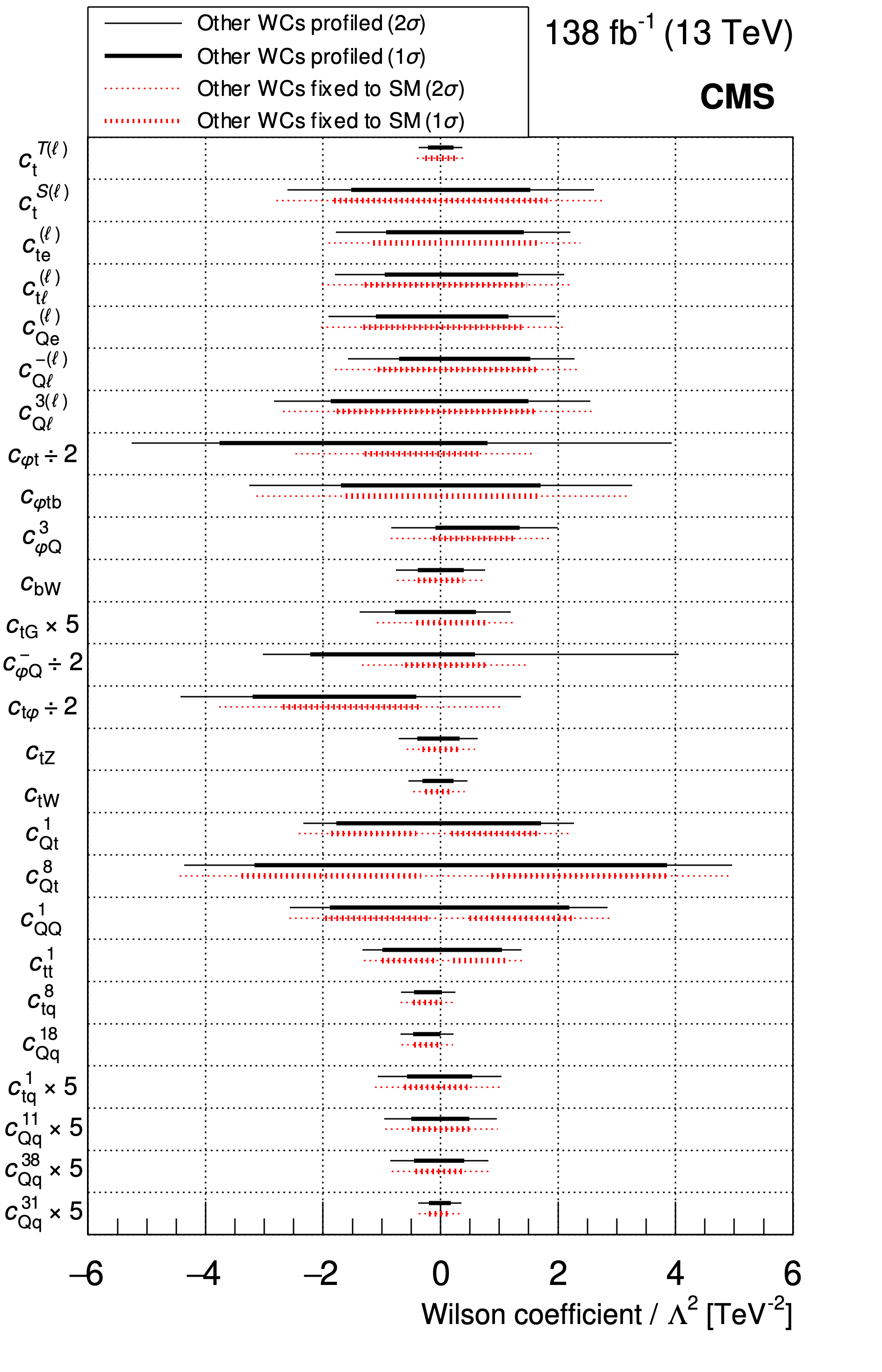}
	\end{minipage} 
	\begin{minipage}{0.55\textwidth}
        \includegraphics[width=0.8\textwidth]{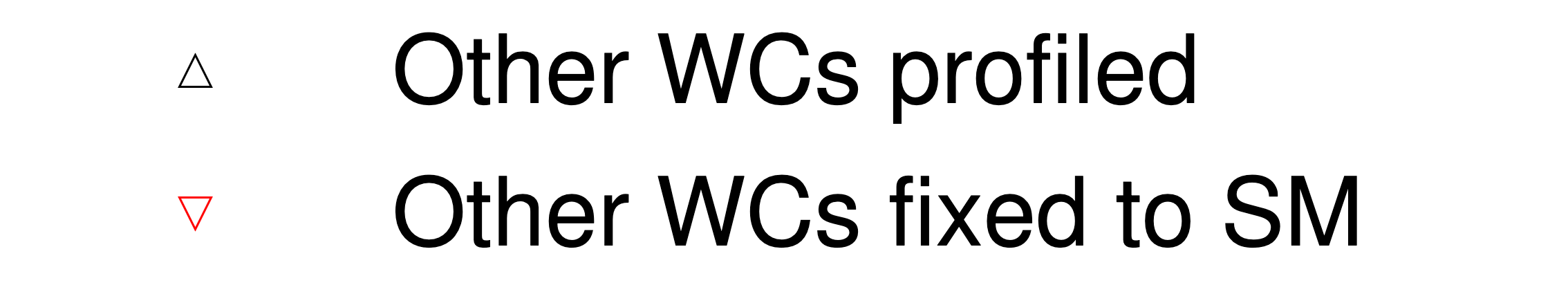} \\
		\includegraphics[width=0.8\textwidth]{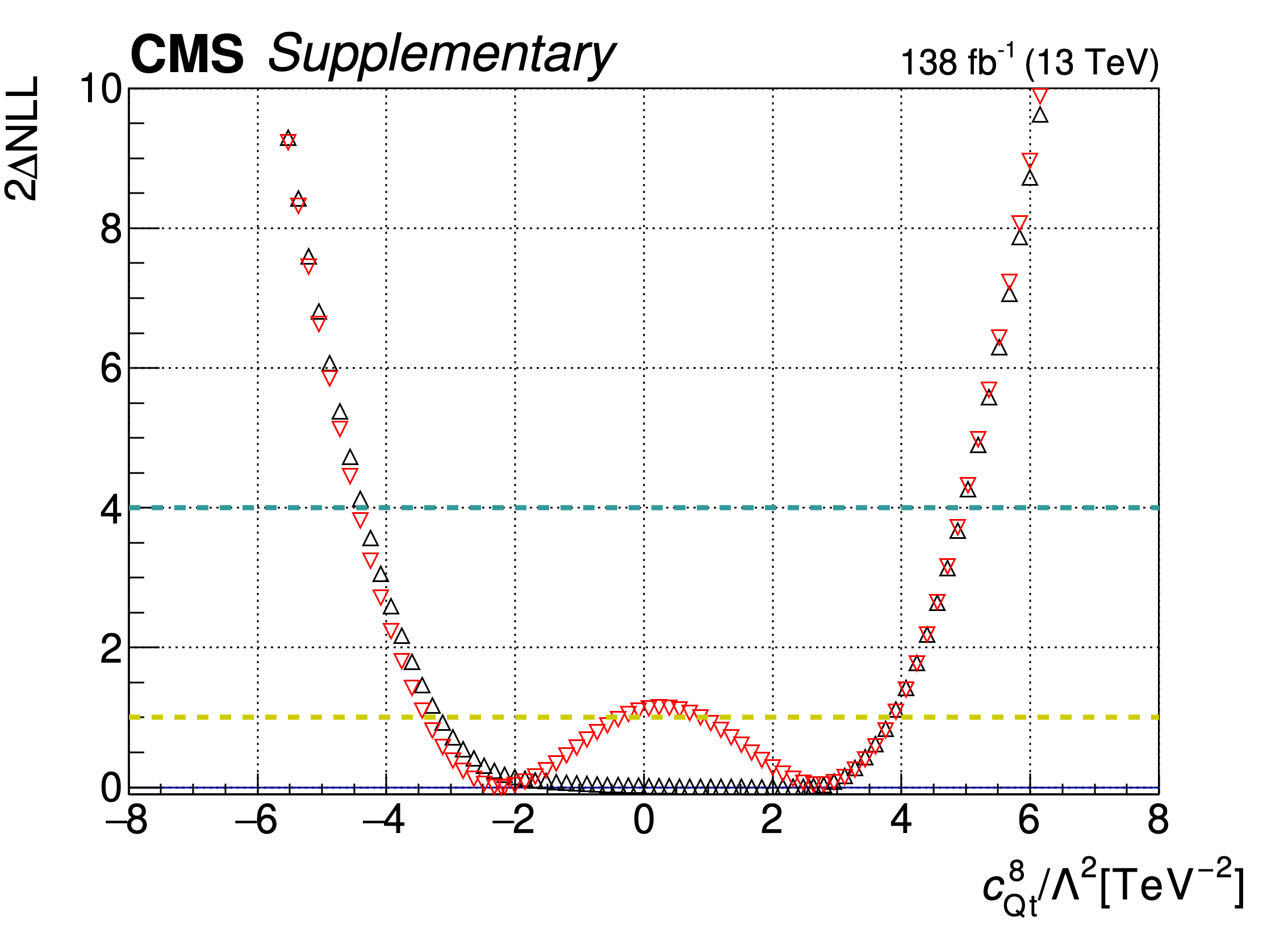} \\
	\end{minipage} 
	\caption{Left: A summary plot showing observed 1$\sigma$ and 2$\sigma$ limits for all the 26 WCs. Right: Observed 2$\Delta$NLL values for 1d scan of \cQtEight. \cite{top22006}}
	\label{fig:1dresult_summarycQt8_1DNLL}
\end{figure}
It is interesting to note that for some WCs such as \cQtEight and \cQQOne, we observe degeneracy in the 1D frozen scan which widens the 1$\sigma$ CIs.

Since the SM expected value of 0 is included at 2$\sigma$ level for all the WCs, no significant deviation from the SM was observed.

\bibliography{eprint}{}
\bibliographystyle{unsrt}
 
\end{document}